# A Framework for Formal Specification and Verification of Security Properties of the Android Permissions System

Amirhosein Sayyadabdi
Faculty of Computer Engineering
University of Isfahan
Isfahan, Iran
ahsa@eng.ui.ac.ir

## ABSTRACT

Android is a widely deployed operating system that employs a permission-based access control model. The Android Permissions System (APS) is responsible for mediating resource requests from applications. APS is a critical component of the Android security mechanism. A failure in the design of APS can potentially lead to vulnerabilities that grant unauthorized access to resources by malicious applications. Researchers have employed formal methods for analyzing the security properties of APS. Since Android is constantly evolving, we intend to design and implement a framework for formal specification and verification of the security properties of APS. In particular, we intend to present a behavioral model of APS that represents the non-binary, context dependent permissions introduced in Android 10 and temporal permissions introduced in Android 11.

## 1 Introduction

Android is the dominant end-user operating sys-tem currently available on the market [1]. Different devices employ Android for managing their hard-ware and software resources and servicing their users, such as smart home appliances, TVs, mobile phones, etc. Android uses the concept of permissions to manage the users' access to resources. Applications can access resources, if and only if they obtain appropriate permissions either via users' consent or declaring permissions in their source code.

The Android Permissions System (APS) is a critical component of Android's security mechanism, and it protects private user data and sensitive sys-tem resources [2]. A flaw in the design or implementation of APS can result in a violation of the security of Android and potentially lead to critical vulnerabilities [3]. Researchers have found permission-related issues in Android, and Google has responded by introducing security patches [3]. Due to the complexity of APS, the patches have not been successful in terminating the vulnerabilities, and the same vulnerabilities with different attack flows remained exploitable [3], [4].

APS evolves as the Android platform evolves, and attackers target its shortcomings to their ad-vantage [2]. Malicious applications can gain unauthorized access to system resources and users' private data because of APS issues.

## 2 Overview of the Proposed Work

Formally specifying the access control mechanism in Android gives a deeper understanding of the operating system, and it allows us to perform a thorough investigation of APS and analyze its security [5].

Prior studies have applied formal methods to security analysis of APS [3], [5], [6]. The distinctive feature of our proposed work is the consideration of temporal permissions and context dependent, non-binary permissions [1] that have not been formally studied yet.

Many APS issues are the design flaws that re-quire "system-wide reasoning" and conventional methods such as testing and static analysis are not very useful in detecting them because those methods are more suited to detect issues in individual components [6].

The purpose of APS is to ensure that unauthorized access to sensitive resources and users' private data is prevented in Android [6]. The main problem that we are trying to address is to validate the security of APS through the formal specification of the behavior of APS and verification of the security properties that APS should satisfy via model checking.

**Contributions:** (1) presenting a model for APS in Android 12, (2) verifying the security properties of APS, (3) design and implementation of a formal verification framework for APS in TLA$^+$.

## 3 Work to be Done

We have four main objectives:
- Presenting an extensible model for APS based on the Attribute-Based Access Control (ABAC) standard.
- Verifying the security properties of the APS model.
- Analyzing the potential vulnerabilities and security flaws of APS.
- Presenting a framework for formal specification and verification of APS.

An overview of our approach is presented in Figure 1. We begin by analyzing the ABAC standard to extract the entities, their relationships, interfaces, and security properties. We then present a basic formal model of ABAC and verify the designated security properties. These steps constitute the first phase of the project. The second phase begins by analyzing the source code of Android 12 along with Google's documentation. We will extract the functionalities and Application Programming Interfaces (APIs) of APS along with a set of required security properties. The basic model of ABAC is then used as a core model in formally specifying the underlying behavioral model of APS. The second phase concludes with verifying the security properties of the model.

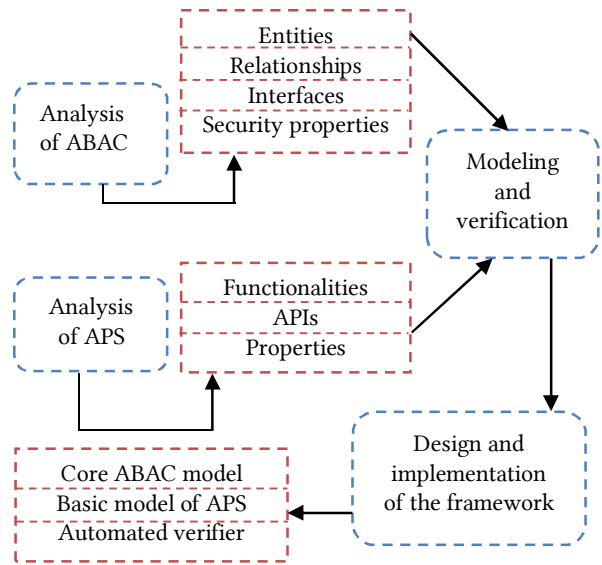

**Figure 1. Overview of the approach**

## 4 Related Work

Formal methods have been applied successfully in the analysis and verification of the securities of APS [3], [5]–[7]. The difference between previous work and this work is three-fold:
- Incorporating the latest developments in APS (e.g., temporal permissions [1]),
- Creating a core model of APS based on ABAC
- Presenting a framework for formal analysis of future versions of APS.

Table 1 presents a brief taxonomy of related work. We categorized the papers based on five criteria: (1) Android version, (2) consideration of the source code in modeling, (3) evaluation method (model checking, proving), (4) consideration of temporal properties, (5) formal language.

**Table 1. Taxonomy of related work**

| Language | Temporal Properties | Model Checking | Proof | Source Code | Android | Paper |
|---|---|---|---|---|---|---|
| - | □ | □ | □ | ■ | 5 | Talegaon et al. [8] |
| Alloy | □ | ■ | ■ | ■ | 6 | Tuncay et al. [3] |
| TLA$^+$ | ■ | ■ | □ | □ | ≥ 6 | Sadeghi et al. [9] |
| Alloy | □ | ■ | □ | □ | 6 | Bagheri et al. [6], [10] |
| TLA$^+$ | □ | ■ | □ | □ | 6 | Sadeghi [11] |
| TLA$^+$ | ■ | ■ | ■ | ■ | 12 | Proposed work |

Legend:    ■ Covered    □ Not covered    - Not stated

## 5 Conclusion

We briefly stated the problem and we noted the differences between our work and related work. We discussed our approach and also taxonomized the core body of related work.

## ACKNOWLEDGMENTS

This work is based upon research funded by Iran National Science Foundation (INSF) under project No 4003042. The author is enrolled in a Ph.D. program advised by Dr. Behrouz Tork Ladani and Dr. Bahman Zamani.